\documentclass[pdflatex,sn-mathphys-num]{sn-jnl}

\usepackage{graphicx}%
\usepackage{multirow}%
\usepackage{amsmath,amssymb,amsfonts}%
\usepackage{amsthm}%
\usepackage{mathrsfs}%
\usepackage[title]{appendix}%
\usepackage{xcolor}%
\usepackage{textcomp}%
\usepackage{manyfoot}%
\usepackage{booktabs}%
\usepackage{algorithm}%
\usepackage{algorithmicx}%
\usepackage{algpseudocode}%
\usepackage{listings}%
\usepackage{url}%

\begin{document}

\title[Phase-space organization of the elastic pendulum]{Phase-space organization of the elastic pendulum: chaotic fraction, energy exchanges, and the order-chaos-order transition}


\author[1]{\fnm{Juan P.} \sur{Tarigo}}\email{juan.tarigo@fcien.edu.uy}

\author[1]{\fnm{Cecilia} \sur{Stari}}\email{cstari@fing.edu.uy}

\author[2]{\fnm{Edson Denis} \sur{Leonel}}\email{edson-denis.leonel@unesp.br}

\author*[1]{\fnm{Arturo C.} \sur{Marti}}\email{marti@fisica.edu.uy}
\affil*[1]{\orgdiv{Instituto de Física}, \orgname{UdelaR}, \orgaddress{\street{Iguá 4225}, \city{Montevideo}, \postcode{11400}, \state{Montevideo}, \country{Uruguay}}}

\affil[2]{\orgdiv{Departamento de F\'isica, Unesp - Universidade Estadual Paulista}, \orgname{UNESP}, \orgaddress{\street{Av.24A, 1515}, \city{Rio Claro}, \postcode{13506-900}, \state{SP}, \country{Brazil}}}

\abstract{
We study the phase-space organization of the planar elastic pendulum
as a function of its two dimensionless control parameters: the reduced
energy $R$ and the squared frequency ratio $\mu$. By randomly sampling the
isoenergetic volume to classify trajectories as oscillatory, rotational,
or chaotic across the $(\mu, R)$ parameter plane, we obtain a global portrait
of the coexistence and competition between dynamical regimes. The chaotic
fraction is not uniformly distributed across the parameter plane but
concentrates in a well-defined central cloud whose ridge follows a linear
relation in the $(\mu, R)$ plane; even where it is most intense,
chaotic motion never fully dominates the accessible phase space, a regular
component always persisting. The order-chaos-order transition is not a global
property of the parameter plane but occurs specifically in the central region
surrounding this cloud: along paths that traverse it, oscillatory orbits
progressively give way to chaotic trajectories, which in turn yield to
rotational orbits as the energy grows, revealing a clear sequential
mechanism underlying the transition. The onset of rotational motion is
gradual rather than sharp, reflecting a strong dependence on initial
conditions. By decomposing the total energy into spring-like,
pendulum-like, and coupling contributions, we establish a direct
correspondence between the coupling power and the abundance of chaotic
trajectories, showing that enhanced inter-mode energy exchange is a
reliable indicator of dynamical complexity. These results provide a
comprehensive and quantitative map of the dynamical regimes of the
elastic pendulum, clarifying the structure of the chaotic cloud and
connecting it to the underlying mode-coupling mechanisms.
}




\maketitle

\section{Introduction}
\label{sec:intro}

In conservative dynamical systems, regular and chaotic trajectories 
generically coexist for the same values of the control parameters, 
depending on initial conditions~\cite{lichtenberg2013regular,
zaslavsky2005hamiltonian}. The manner in which they coexist, the structure 
of phase space, the way they intermingle, and their relative proportions 
constitute a rich and complex problem that remains far from fully 
understood. This complexity stems from the fact that a generic Hamiltonian 
system is neither integrable nor ergodic: it lives in the vast intermediate 
territory between these two extremes, where regular islands and chaotic 
seas coexist in an intricate, parameter-dependent 
arrangement~\cite{markus1974generic}. The prevailing approach to this 
problem --- one that dates back to the very origins of analytical 
mechanics --- consists of starting from a known integrable Hamiltonian and 
introducing a small perturbation, thereby allowing the use of perturbation 
theory~\cite{Goldstein,zaslavsky2005hamiltonian,lemos2018analytical} to 
track the gradual breakdown of regular motion.

Within this framework, a recurring and well-documented phenomenon is the 
order-chaos-order transition: periodic or quasiperiodic solutions 
predominate at low energies, chaotic solutions emerge and grow at 
intermediate energies without occupying the entire available phase space, 
and regular trajectories regain prevalence at higher energies due to the 
restoration of symmetries. Reichl and Brumer~\cite{reichl1986stochastic} 
identified this transition in a system of harmonically coupled oscillators 
with a nonlinear coupling function, while Bolotin et 
al.~\cite{bolotin1989transition} demonstrated it in a generalization of 
the celebrated Hénon-Heiles potential. Notably, Deng and 
Hioe~\cite{deng1985chaos} studied two nonlinearly coupled oscillators and 
found not only the order-chaos-order sequence but also the existence of 
additional parameter windows where the dynamics abruptly regularizes --- a 
particularly striking result, as it implies multiple chaos-order transitions 
in a system that is proven to be non-integrable for generic parameter values \cite{goriely2001integrability}.

The statistical characterization of chaos across phase
space---through the fraction of trajectories that are chaotic rather than
regular---has recently attracted attention in a variety of low-dimensional
systems. In a previous work~\cite{cabrera2025regular}, we showed that the
double pendulum undergoes an order--chaos--order transition as a function of
mechanical energy, quantified via Poincaré sections and maximum Lyapunov
exponent calculations, with the chaotic fraction growing exponentially in the
low-energy regime. This question was subsequently taken up by
Jiménez-López and García-Garrido~\cite{jimenez2024chaos}, who used Lagrangian
descriptors to quantify the chaotic fraction of phase space as a function of
the total energy and of the masses, lengths, and gravity, finding that, for a
fixed mass ratio, this fraction is maximized when both pendulums have equal
lengths, and exploring the hypothesis that it grows and decays following an
exponential law across different energy regimes. Analogous behavior occurs in
the transition from integrability to chaos in an oval billiard driven by
boundary deformation, identifying an order parameter whose saturation value
follows a scaling law analogous to that of second-order phase transitions in
statistical mechanics \cite{leonel2026describing}. More recently, Sander and Meiss~\cite{sander2025proportions}
studied dynamical behavior classification in one- and two-dimensional torus
maps, showing that while Arnold's circle map obeys a universal power law for
the fraction of nonresonant orbits, no such universality holds in the
two-dimensional case. In a recent study~\cite{moraes2026framework}, it was shown that rigid choices of coordinates and of the Poincar\'e sectioning hyperplane can introduce geometric distortions and coordinate artifacts that obscure or clip fundamental invariant structures; the identification of the different types of trajectories therefore depends on the section adopted, and at least two complementary, field-conforming sections may be required to obtain a distortion-free diagnostic of the phase-space organization.
For this reason, a methodological contribution
of the present work is to sample initial conditions randomly within the
energetically accessible three-dimensional phase space, thereby avoiding the
biases associated with any particular section or choice of a particular Poincaré section. Another recent example is the work~\cite{benvenisty2025elastic}, which introduces a new category of response in the planar elastic pendulum: a slow,
sequential back-and-forth switching between a nearly periodic motion and its
mirror image, reminiscent of the Dzhanibekov effect.

In the present work, we study the phase-space organization of the elastic
pendulum, a paradigmatic system whose intrinsic nonlinearity and
autoparametric resonance preserve several generic features of nonlinear
systems while also exhibiting unusual and system-specific characteristics.
Our main contribution is a global, statistical characterization
of its dynamics across the entire $(\mu, R)$ parameter plane, rather than the
description of individual orbits, which can be summarized as follows.
We address the relative abundance of regular and chaotic orbits as a function
of its two control parameters, showing that regular orbits can be classified as
oscillatory or rotational, with parameter-dependent boundaries and a transition
region where both types coexist; as the energy is increased, the
system displays an order--chaos--order sequence, evolving from predominantly
oscillatory motion, through a chaotic regime, to rotational motion. Building on
the energy decomposition framework of de Souza et
al.~\cite{de2018energy,de2022internal}, we compute each energy component and its
associated power, and use the energy exchanges between the
elastic and pendular modes as an indicator of dynamical complexity, showing how
they relate to the abundance of chaotic trajectories. Finally,
we find that chaotic orbits organize themselves into a well-defined cloud in
parameter space where the maximum chaotic fraction does not exceed $70\%$,
flanked by regions where each type of regular trajectory
predominates, with the locus of maximal chaoticity following a
simple linear relation in the $(\mu, R)$ plane.

The remainder of the paper is organized as follows.
Section~\ref{sec:pep} introduces the elastic pendulum, reviews prior
contributions, and presents the equations of motion in
dimensionless form, identifying the two control parameters used throughout: the
reduced energy $R$ and the squared frequency ratio $\mu$.
Section~\ref{sec:res} presents the main results: we first
describe how trajectories are sampled within the energetically accessible volume
and classified, by means of Poincaré sections, as oscillatory, rotational, or
chaotic; we then map the relative abundance of each regime across the
$(\mu, R)$ plane, characterize the chaotic cloud and the linear ridge along
which chaos is most intense, and analyze the order--chaos--order transition;
finally, we decompose the energy into spring-like, pendulum-like, and coupling
contributions and relate the coupling power to the abundance of chaos.
Conclusions and directions for future work are drawn in
Section~\ref{sec:con}.

\section{The Elastic Pendulum}
\label{sec:pep}

This section presents the model, its Hamiltonian formulation, and the 
numerical methods employed. We pay particular attention to the 
implementation details of the latter, which are often delicate and 
frequently omitted in the literature despite their significant impact 
on the reliability of the results.

\subsection{Physical description and dynamical equations}

\begin{figure}[ht]
\centering
\includegraphics[width=0.5\textwidth]{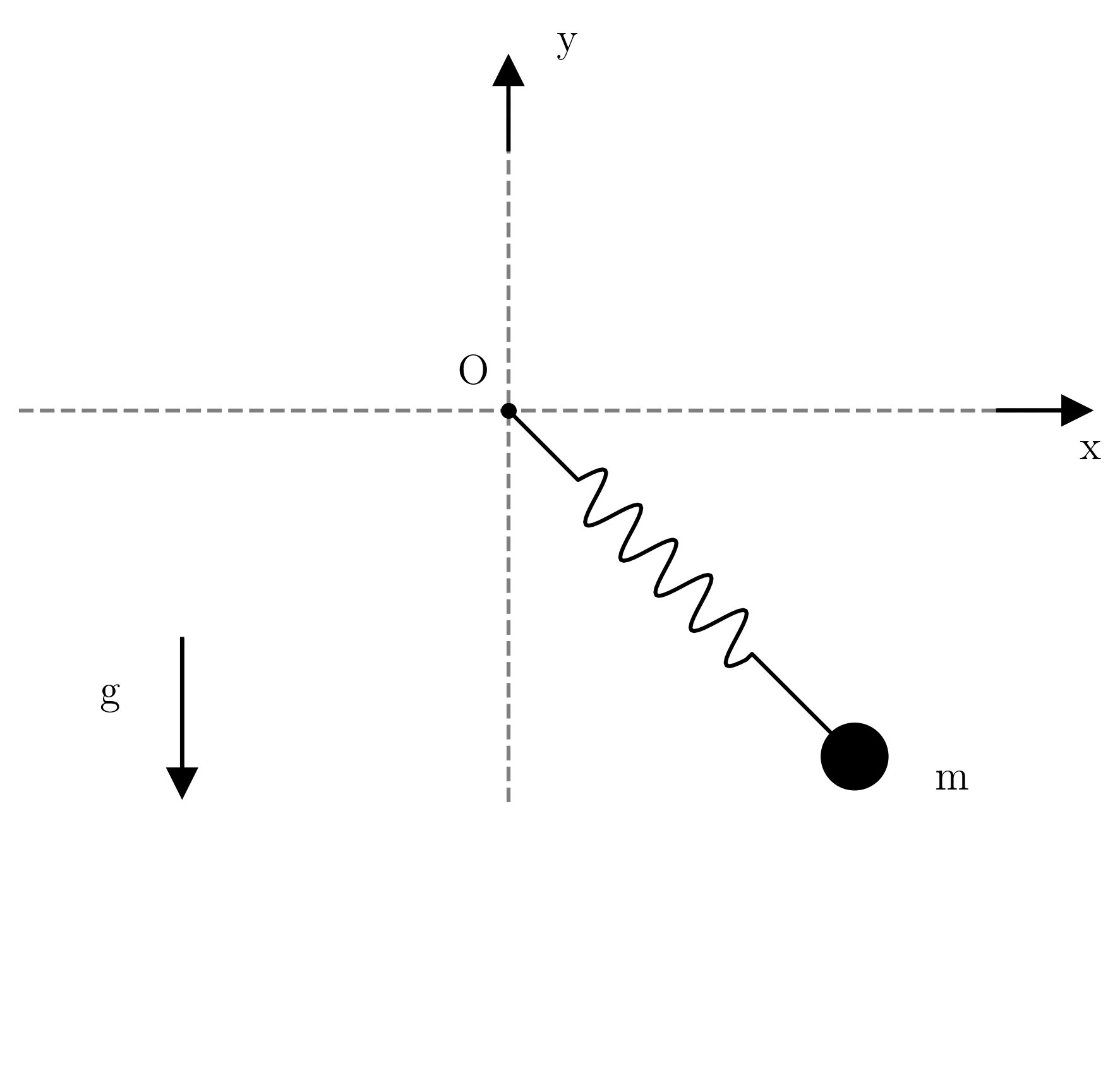}
\caption{The planar elastic pendulum composed of a spring of natural length 
$l_0$ and stiffness $k$. Throughout this paper Cartesian canonical coordinates 
are used.}
\label{fig1}
\end{figure}

The elastic (or spring) pendulum is obtained by replacing the inextensible 
rod of a simple pendulum with an ideal spring.
Specifically, it  consists of a point mass $m$ 
attached to a spring of constant $k$ and natural length $l_0$, subject to 
a gravitational field $g$, as shown in Fig.~\ref{fig1}. We consider the 
planar case, in which the motion is restricted to a vertical plane, so that 
the system has two degrees of freedom. In the absence of friction, which is 
the standard assumption in the study of its conservative 
dynamics~\cite{cuerno1992deterministic,carretero1994regular,nunez1990onset,
van1996order,de2018energy,anurag2022locating}, the total mechanical energy 
is the only conserved quantity. Despite its apparent simplicity, this 
two-degree-of-freedom Hamiltonian system exhibits a remarkably rich 
dynamical behavior that has attracted sustained interest over several 
decades.

Its linearized dynamics is characterized by the superposition of two normal 
modes: an elastic (stretching) mode with frequency $\omega_s = \sqrt{k/m}$ and 
a pendular mode with frequency $\omega_p = \sqrt{g/l_e}$, where 
\begin{equation}
    l_e = l_0 + \frac{mg}{k}
\end{equation}
is the equilibrium length of the spring. A particularly prominent feature of the 
system is the autoparametric resonance that occurs when the elastic mode frequency 
is twice the pendular frequency, a condition intrinsic to the system itself rather 
than imposed by an external drive. Near this resonance, rapid and sustained energy 
exchange between the two modes takes place, acting as a primary mechanism for the 
onset of chaotic motion~\cite{cuerno1992deterministic,carretero1994regular,
nunez1990onset}.

The pioneering works of Cuerno \textit{et al.}~\cite{cuerno1992deterministic}, 
Carretero-González \textit{et al.}\ and Núñez-Yépez \textit{et 
al.}~\cite{carretero1994regular,nunez1990onset} documented a remarkable 
alternation between regular and chaotic regimes as a function of the total 
energy: regular behavior at low energies, predominantly chaotic dynamics at 
intermediate energies, and a return to regular behavior at high energies. This 
order-chaos-order transition was studied systematically by Van der Weele and de 
Kleine~\cite{van1996order}, who mapped the dynamics as a function of two 
dimensionless control parameters: the reduced energy $R$ and the squared 
frequency ratio $\mu = \omega_s^2/\omega_p^2$. Their analysis revealed that 
regular behavior predominates in the limiting regions of this parameter plane, 
while chaotic motion is concentrated in a central ``chaotic cloud,'' which they 
interpreted in terms of the competition
between oscillating and rotating trajectories. 
Anurag \textit{et al.}~\cite{anurag2020understanding} revisited this transition 
using a truncated coupling model and a Chirikov overlap criterion, later extended 
to the three-dimensional elastic pendulum~\cite{anurag2022locating}. The elastic 
pendulum has also motivated extensions to the dissipative 
setting~\cite{alasty2006chaotic}, where periodic forcing produces fractal basins 
of attraction. 

In more recent papers, de Souza et al.~\cite{de2018energy,de2022internal} proposed a
decomposition of the total energy into spring-like, pendulum-like, and coupling 
contributions, showing that energy exchange rates are significantly enhanced for 
chaotic orbits, with the highest coupling activity occurring near the homoclinic 
tangle of the primary hyperbolic point. Finally, Acosta-Zamora et 
al.~\cite{acosta2024characterization} studied the geometry of phase-space 
trajectories using Poincaré maps, showing that regular orbits are closely related 
to torus knots and can be characterized by a rational parameter naturally described 
by Farey sequences.

Using Cartesian coordinates $(x, y)$ for the position of the mass, with $y$ 
measured upward from the suspension point, the Lagrangian of the system is
\begin{equation}
    L = \frac{1}{2}m \left(\dot{x}^2 + \dot{y}^2\right) - mgy
        - \frac{1}{2}k \left( \sqrt{x^2+y^2} - l_0 \right)^2,
\end{equation}
and the total mechanical energy, which is conserved, reads
\begin{equation}
    E = \frac{1}{2}m \left(\dot{x}^2 + \dot{y}^2\right) + mgy
        + \frac{1}{2}k \left( \sqrt{x^2+y^2} - l_0 \right)^2.
\end{equation}
The minimum energy corresponds to the mass at rest at the equilibrium position 
$(0, -l_e)$,
\begin{equation}
   E_{\min} = -mg \left( l_0 + \frac{1}{2}\frac{mg}{k} \right),
\end{equation}
which allows us to introduce the dimensionless energy parameter
\begin{equation}
    R \equiv -\frac{E}{E_{\min}},
\end{equation}
so that $R = -1$ at equilibrium and $R > -1$ for any excited state. Together with 
the frequency ratio
\begin{equation}
    \mu \equiv \frac{\omega_s^2}{\omega_p^2} = 1 + \frac{k l_0}{mg},
\end{equation}
the pair $(\mu, R)$ constitutes the two dimensionless control parameters of the 
problem~\cite{van1996order}. Note that $\mu > 1$ for all physically admissible 
configurations, and the autoparametric resonance condition corresponds to $\mu = 4$.

\subsection{Dimensionless Hamiltonian formulation}

The Hamiltonian formulation is particularly well suited for the study of 
conservative dynamics: by Liouville's theorem, the flow in phase space 
preserves volume, and the freedom to choose the most convenient set of 
canonical variables allows for a natural nondimensionalization of the 
equations of motion \cite{Goldstein,lemos2018analytical}. Introducing dimensionless coordinates and momenta,
\begin{equation}
    q_1 = \frac{x}{l_0}, \quad q_2 = \frac{y}{l_0}, \quad
    p_1 = \frac{\dot{x}}{\sqrt{g l_0}}, \quad p_2 = \frac{\dot{y}}{\sqrt{g l_0}},
\end{equation}
and dividing the energy by $mgl_0$, the dimensionless Hamiltonian takes the 
compact form
\begin{equation}
    \mathcal{H}(q_1, q_2, p_1, p_2) = \frac{1}{2} \left(p_1^2 + p_2^2\right) 
    + q_2 + \frac{1}{2}(\mu - 1) \left( \sqrt{q_1^2 + q_2^2} - 1 \right)^2.
    \label{eq:hamiltonian}
\end{equation}
The nonlinear coupling between the two degrees of freedom is entirely encoded in 
the factor $\left(1 - 1/\sqrt{q_1^2+q_2^2}\right)$, which vanishes in the linear 
regime. Hamilton's equations of motion read
\begin{equation}
\left\{
\begin{aligned}
    \dot{q}_1 &= p_1, \\[4pt]
    \dot{p}_1 &= -(\mu - 1)\, q_1 \left( 1 - \frac{1}{\sqrt{q_1^2 + q_2^2}} 
                \right), \\[4pt]
    \dot{q}_2 &= p_2, \\[4pt]
    \dot{p}_2 &= -1 - (\mu - 1)\, q_2 \left( 1 - \frac{1}{\sqrt{q_1^2 + q_2^2}} 
                \right),
\end{aligned}
\right.
\label{eq:eom}
\end{equation}
which constitute the system we integrate numerically throughout this work. 
The value of $\mathcal{H}$ is fixed by the initial conditions and relates to 
the control parameters through
\begin{equation}
    \mathcal{H} = R\,\frac{2\mu - 1}{2(\mu - 1)},
    \label{eq:H_Rmu}
\end{equation}
as follows directly from the definitions of $R$, $E_{\min}$, and $\mu$.

\subsection{Numerical integration, trajectory classification, and Poincaré surface of section}

Equations~\eqref{eq:eom} are integrated using a ninth-order explicit Runge-Kutta
algorithm with an eighth-order interpolant~\cite{verner2010numerically}.
Energy conservation is monitored via Eq.~\eqref{eq:H_Rmu} as a strict precision criterion
at each integration step: if the relative energy error exceeds a prescribed tolerance,
the integration step is repeated with a reduced step size until the target precision is achieved. 
With this method, the pointwise relative energy error remains below $10^{-10}$ throughout
each trajectory, and the cumulative root-mean-square energy error, accumulated over all
integration steps, does not exceed $10^{-6}$.

To calculate the statistical fractions of the phase-space dynamics, an ensemble of $10,000$ initial 
conditions are randomly sampled for each parameter pair. These initial conditions are 
chosen uniformly inside the isoenergetic volume. Specifically, the coordinates $q_1$, $q_2$, 
and the momentum $p_2$ are sampled randomly, while the remaining momentum $p_1$ is uniquely 
determined from the energy constraint under the condition $p_1 > 0$.

Each integrated trajectory is then classified as either chaotic or regular using the 
Generalized Alignment Index of third order ($\text{GALI}_3$) \cite{skokos2007generalized}. 
For trajectories identified as regular, we further classify their bounded motion as either 
rotational or oscillating. This is done by tracking the polar angle in the configuration
space; if the trajectory completes a full $2\pi$ rotation in this angle, it is classified
as circular, otherwise it is classified as oscillating.

The phase-space structure is further visualized through a Poincaré surface of section, 
defined by the condition $q_1 = 0$ and $p_1 > 0$. On this hyperplane, Eq.~\eqref{eq:H_Rmu} 
determines $p_1$ uniquely as 
\begin{equation}
    p_1 = +\sqrt{ \frac{R(2\mu - 1)}{\mu - 1} - p_2^2 - 2q_2 - (\mu - 1) \left( |q_2| - 1 \right)^2 } \geq 0\text{,} \label{eq:sos_constraint}
\end{equation}
so that each trajectory is recorded as a point $(q_2, p_2)$ each time it crosses this hyperplane in the prescribed direction. This construction reduces the four-dimensional phase space to a two-dimensional map, providing an efficient visual diagnostic to complement the $\text{GALI}_3$ classification, cleanly distinguishing regular trajectories—which trace out closed curves associated with KAM tori—from chaotic ones, which fill two-dimensional regions ergodically.

\section{Results}\label{sec:res}

We present the results in two complementary parts. We first characterize 
the geometry of phase space by classifying trajectories and mapping their 
relative abundance across the control parameter plane 
(Section~\ref{sec:portrait}). We then connect this global portrait to the 
underlying physical mechanism by analyzing the distribution of energy 
among the elastic, pendular, and coupling modes and the rates at which 
they exchange it (Section~\ref{sec:energy}).

\subsection{Phase-space portrait and trajectory classification}
\label{sec:portrait}

As discussed in Section~\ref{sec:pep}, the planar elastic pendulum is a 
two-degree-of-freedom system with a single conserved quantity, the energy, 
so that the dynamics is confined to a three-dimensional energy surface in 
phase space. The two control parameters are the dimensionless energy $R$ and 
the frequency ratio $\mu$. For a given pair $(R, \mu)$, different types of 
trajectories generally coexist depending on initial conditions. A first 
qualitative appreciation of this diversity can be gained by inspecting 
trajectories directly in configuration space which, while not fully 
determining the dynamics, provides an intuitive picture of the possible 
orbital morphologies.

\begin{figure}[hbpt]
\centering
\includegraphics[width=0.80\textwidth]{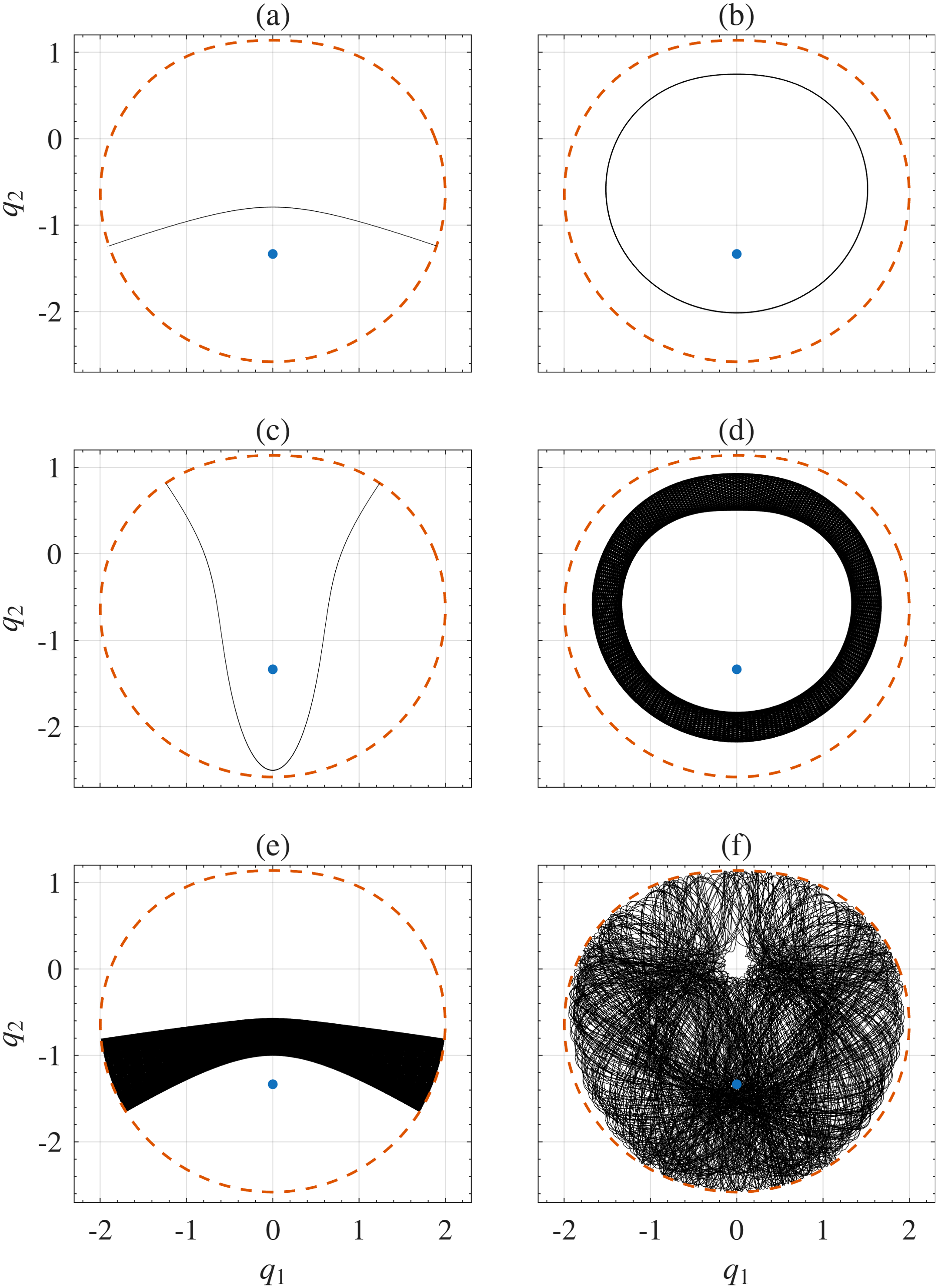}
\caption{Example of trajectories in configuration space $(q_1, q_2)$. Panel (a), (c) and (e) illustrates oscillations, (b) and (e) show a rotation trajectories where the pendulum completes a full circle. Panel (d) exhibits an example of a chaotic trajectory. Parameters are $\mu = 4$ and $R = 1$ while initial conditions are $q_1 = p_2 = 0$ and $q_2 =$ $-0.7905$ for (a), $0.75$ for (b), $-2.505$ for (c),  $0.5$ for (d),  $-1.0$ for (e)  $-1.5$ for (f).}\label{fig:trajectories}
\end{figure}

Six representative examples are shown in Fig.~\ref{fig:trajectories}. 
Panels (a) and (c) display oscillatory trajectories in which the mass does 
not complete a full revolution around the suspension point; these correspond 
to what Van der Weele and de Kleine~\cite{van1996order} classified as 
``hill'' and ``valley'' orbits, respectively, both belonging to the 
oscillatory orbit family. Panel (e) shows similar dynamics to (a) and 
(c) but for a quasi-periodic trajectory. Panel (b) shows a rotational 
trajectory, in which the mass completes full revolutions, panel (d) 
shows a quasi-periodic example of this dynamics. Finally, panel (f) 
shows a chaotic trajectory that progressively fills the available 
region of configuration space as time evolves, with no apparent spatial regularity.

In this work, we study the relative abundance of the different trajectory 
types present in the elastic pendulum and how they intermingle to form a 
characteristic entangled structure that depends strongly on the control 
parameters. Within the regular orbits, we distinguish between oscillatory 
and rotational trajectories and examine how their distribution across phase 
space influences the abundance of chaos.

\begin{figure}[htbp]
\centering
\includegraphics[width=0.75\textwidth]{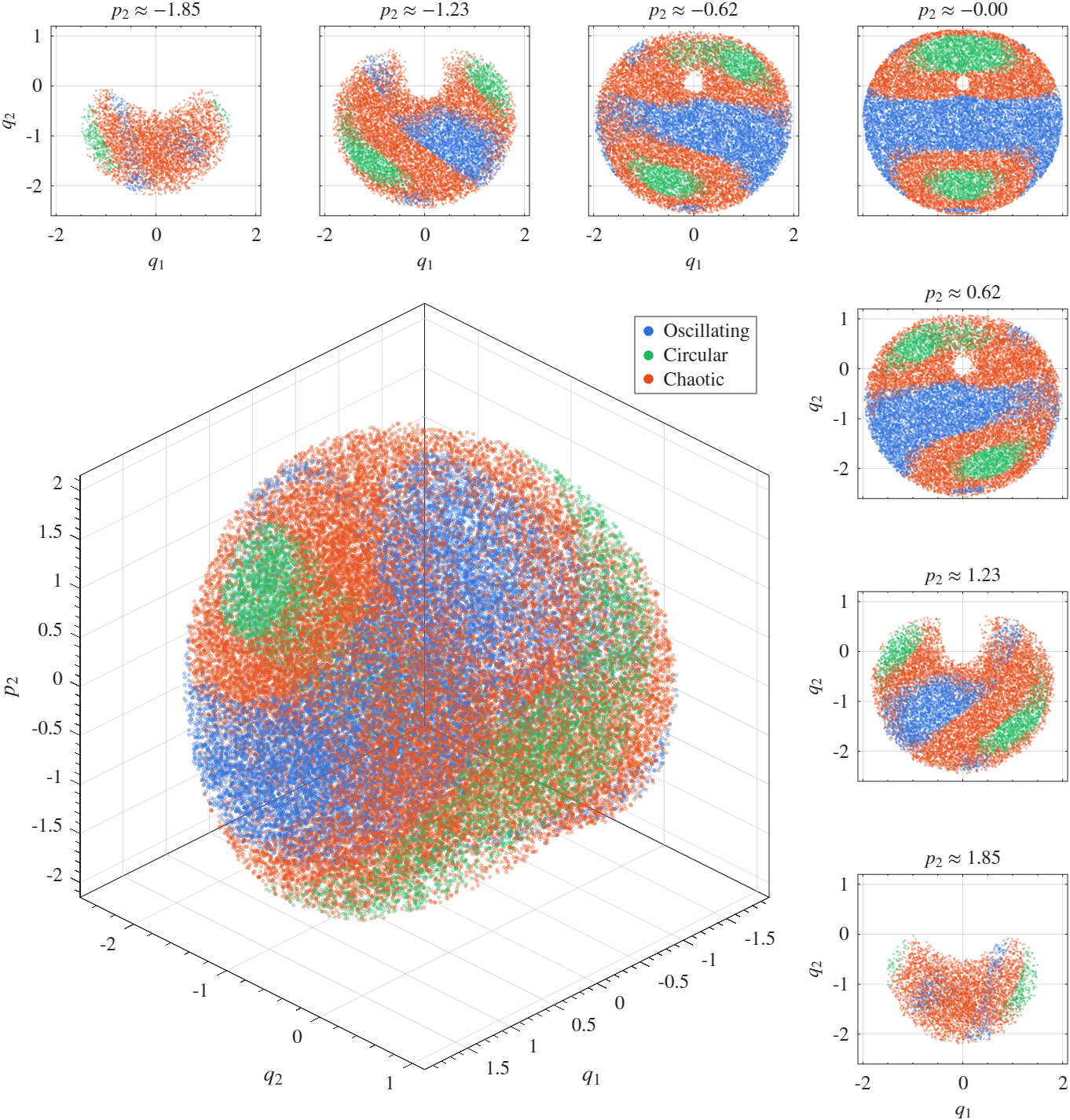}
\caption{Three-dimensional visualization of the phase space volume for parameters $\mu = 4$ and $R = 1$ in the $(q_1, q_2, p_2)$ subspace. The central 3D scatter plot illustrates the spatial distribution of oscillating (blue), circular/rotating (green), and chaotic (red) trajectories. The surrounding two-dimensional panels display horizontal cross-sections of this volume at various values of $p_2$, highlighting the internal structure and boundaries of the different dynamics.}\label{fig:volume}
\end{figure}

Figure~\ref{fig:volume} presents a visualization of the three-dimensional 
isoenergetic surface together with seven cross-sections at constant values of 
$q_1$, in which oscillatory, rotational, and chaotic trajectories can be 
distinguished. Oscillatory trajectories tend to cluster in a region around 
the origin, while rotational trajectories appear grouped in two disjoint 
regions on either side; in both cases these regular islands are surrounded by 
a chaotic sea, shown in red.

\begin{figure}[htbp]
\centering
\includegraphics[width=0.80\textwidth]{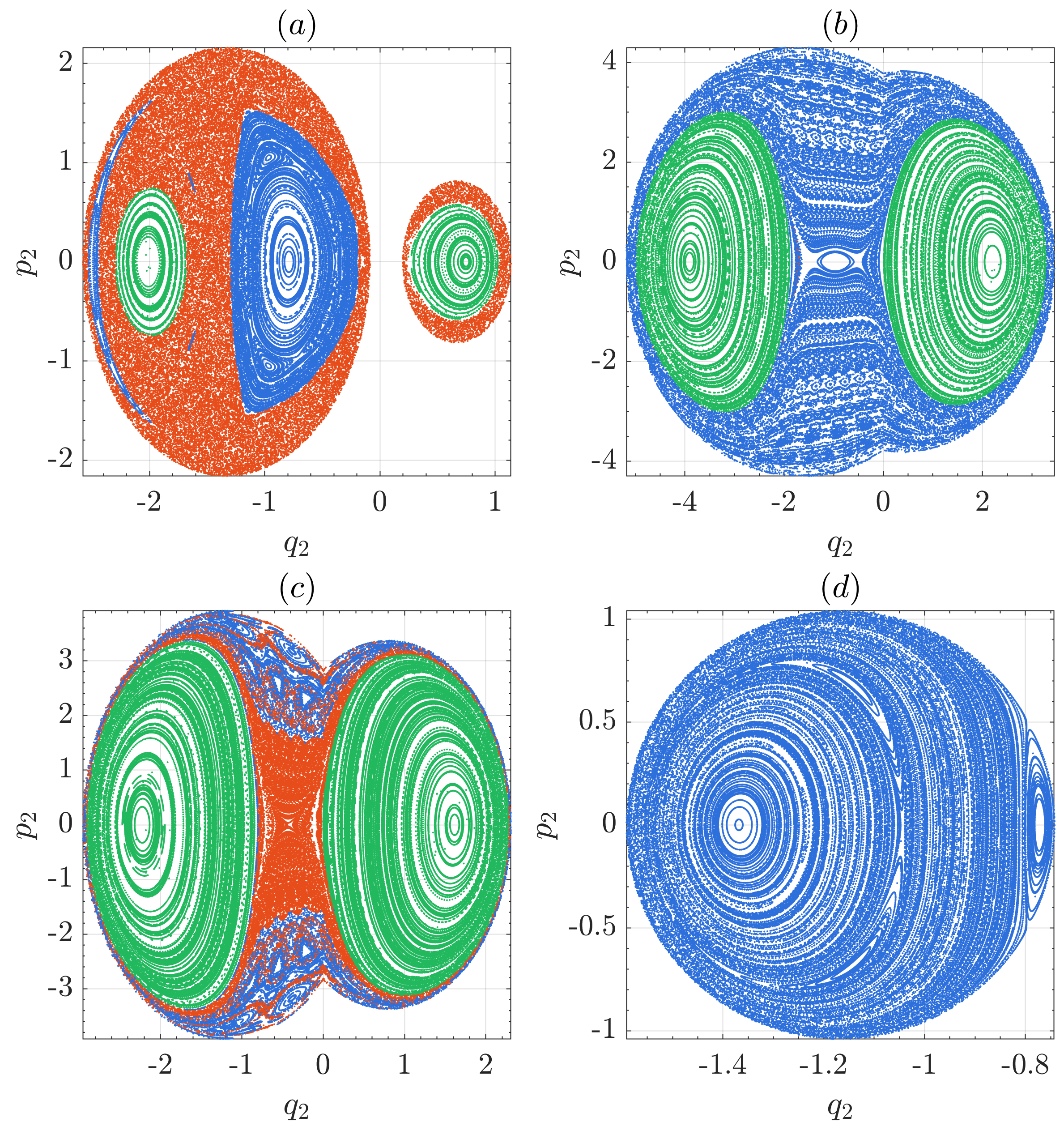}
\caption{Poincaré sections at the plane $q_1 = 0$ with $p_1 > 0$, illustrating the phase space structure for different parameter combinations. The panels correspond to $(\mu, R)$ values of: (a) $(4, 1)$, (b) $(2.5, 6)$, (c) $(6, 6)$, and (d) $(7, -0.5)$. The diagrams reveal the coexistence of regular regions (invariant tori corresponding to oscillating or rotating motion in blue or green respectively) and chaotic seas (in red), which vary significantly depending on the chosen parameters.}\label{fig:psections}
\end{figure}

The phase-space structure is further illustrated through Poincaré surfaces of 
section in Fig.~\ref{fig:psections} for four representative pairs of control 
parameters. Panel (a) corresponds to the same parameter values shown in 
Fig.~\ref{fig:volume}. Panels (a) and (c) display situations where all three 
trajectory types coexist, panel (b) shows both types of regular trajectories 
in the absence of significant chaos, and panel (d) contains exclusively 
oscillatory orbits.

It is worth noting the distinction between Figs.~\ref{fig:volume} and 
\ref{fig:psections}: while both represent cross-cut through phase space, the former 
shows slices of finite thickness at several values of $q_1$, whereas the 
Poincaré sections record intersection points with a single, precisely defined 
hyperplane ($q_1 = 0$, $p_1 > 0$), yielding a strict two-dimensional portrait 
of the dynamics.

\begin{figure}[ht]
\centering
\includegraphics[width=0.99\textwidth]{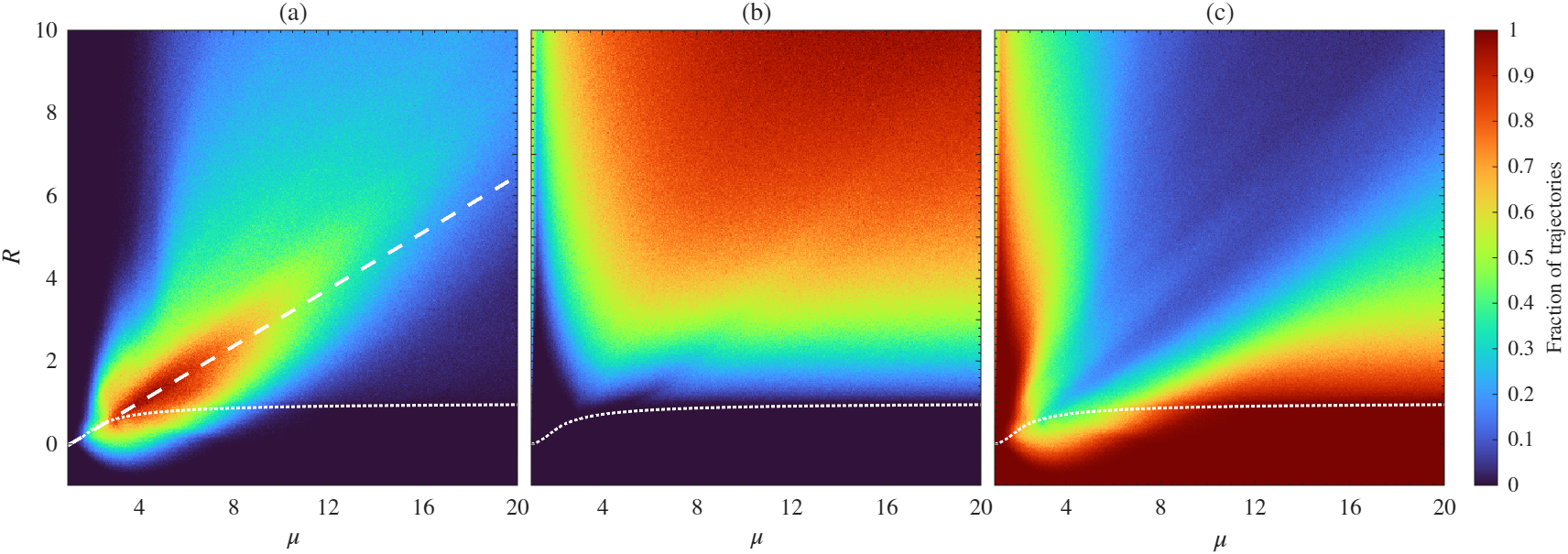}
\caption{Fraction of chaotic [panel~(a)], rotational [panel~(b)],
and oscillatory [panel~(c)] trajectories across the $(\mu, R)$ parameter plane.
Each panel shows the fraction of trajectories of a given type, estimated from
$10^4$ initial conditions sampled randomly within the isoenergetic volume for
each parameter pair; the color scale ranges from $0$ to $1$. The dotted line
marks the minimum energy threshold for rotational motion, given by
Eq.~\eqref{eq:rotation_energy}. The dashed line, shown in
panel~(a), indicates the location of maximum chaotic fraction in the parameter
plane, whose linear fit is discussed in the text, Eq.~\eqref{eq:maximums}.}
\label{fig:fractions}
\end{figure}

In order to quantify the distribution of trajectory types across the parameter
space, Fig.~\ref{fig:fractions} shows the relative fractions of
oscillatory   [panel~(c)], rotational  [panel~(b)],
and chaotic  [panel~(a)] trajectories as a function
of the control parameters  $(\mu, R)$.
For each parameter pair, the fractions were estimated from $10^4$
trajectories whose initial conditions were sampled randomly inside the
isoenergetic volume defined by Eq.~\eqref{eq:sos_constraint}; by construction,
the three fractions add up to one at every point of the plane.

At low energies ($R \approx -1$), the vast majority of trajectories are
oscillatory. As the energy increases, chaotic trajectories  first
emerge and grow in abundance, before eventually giving way to
rotational orbits  at the highest energies.
This order--chaos--order progression as the energy is raised is
consistent with the scenario reported  in
Ref.~\cite{van1996order}.
The minimum energy required for rotational motion to be kinematically possible is
given by~\cite{van1996order}:
\begin{align}
    R &= \frac{(\mu-1)}{2+1/(\mu-1)} \quad \text{for} \quad 1<\mu<2,\\
    R &= 1 - \frac{2}{2(\mu-1)+1} \quad \text{for} \quad \mu>2.
    \label{eq:rotation_energy}
\end{align}
This threshold is indicated by the dotted line in Fig.~\ref{fig:fractions}.
However, this condition is necessary but not sufficient: even when the energy
exceeds this threshold, whether a given trajectory is rotational or not depends
on the remaining initial conditions. This is reflected in the gradual rather than
sharp onset of rotational orbits seen in panel~(b), where
oscillatory and rotational trajectories coexist over a broad region of the
parameter plane above the threshold defined by Eq.~\eqref{eq:rotation_energy}.
In the region $\mu \to 1$ we also observe a predominance of oscillatory
orbits. Since $\mu$ measures the squared ratio of the elastic and pendular
frequencies, the $2{:}1$ (autoparametric) resonance occurs at $\mu = 4$,
so that the limit $\mu \to 1$ corresponds to comparable frequencies
($\omega_s \approx \omega_p$) and thus lies far from resonance. In this
regime the coupling between the elastic and pendular modes is weak, the two
modes barely exchange energy, and the motion remains essentially regular,
dominated by oscillatory orbits.

Panel~(a) further shows that chaotic trajectories are not
uniformly distributed across the parameter plane but instead concentrate in a
well-defined central region, whose maximum traces a ridge
oriented obliquely in the $(\mu, R)$ plane. To characterize this
structure more precisely, we fitted a linear function to the locus of maximum
chaotic fraction for each value of $\mu$, obtaining:
\begin{equation}
    R_{\max} = 0.34\,\mu - 0.38,
    \label{eq:maximums}
\end{equation}
which is represented by the dashed line in Fig.~\ref{fig:fractions}.
The positive slope indicates that progressively higher energies
are needed to reach the most strongly chaotic regime as $\mu$ increases.

\begin{figure}[ht]
\centering
\includegraphics[width=0.50\textwidth]{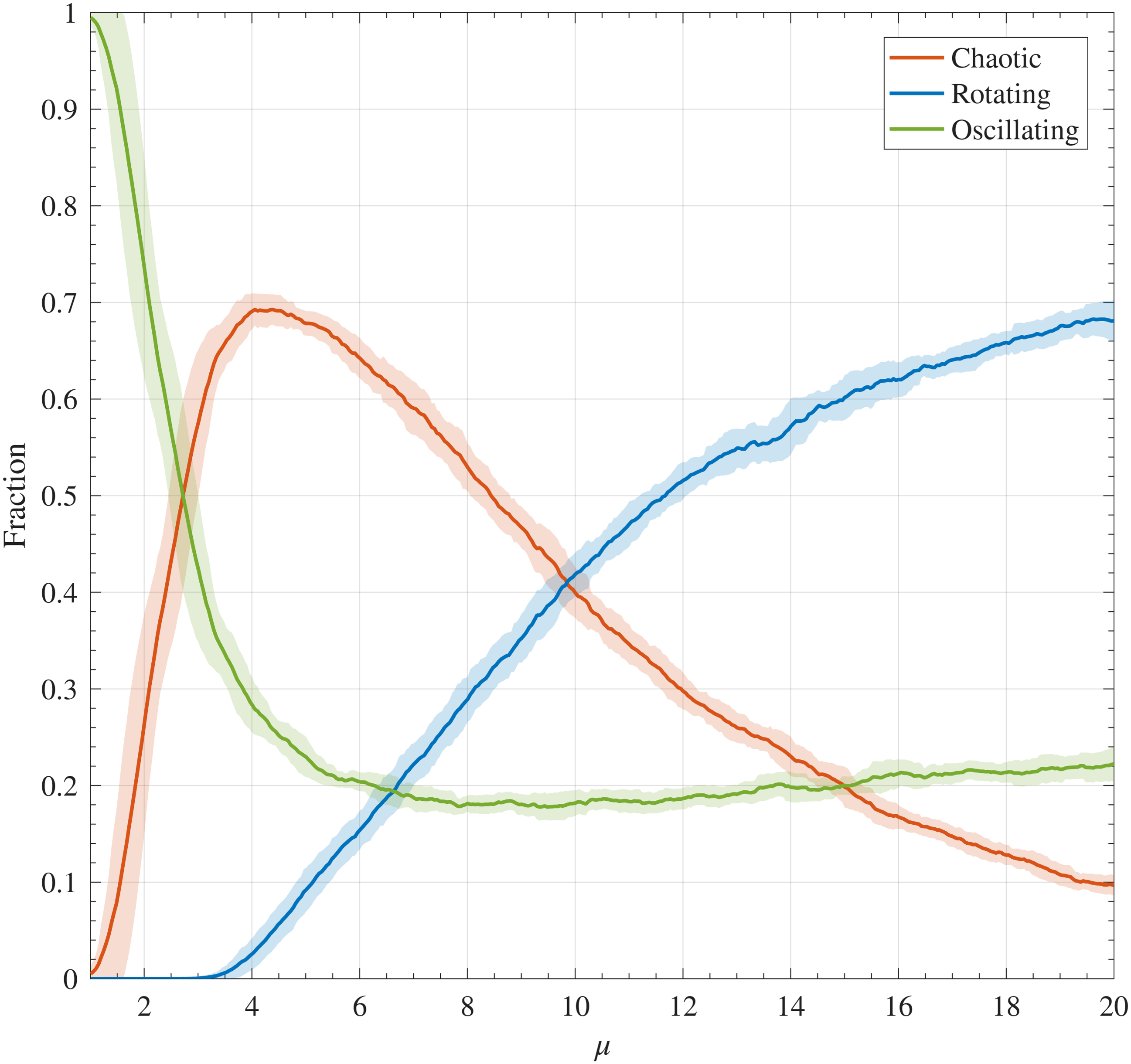}
\caption{Trajectory type fractions along the linear ridge of maximum
  chaotic fraction defined by Eq.~\eqref{eq:maximums}. The solid lines
  show the moving averages of the oscillatory (blue), rotational
  (green), and chaotic (red) fractions sampled along this
  straight-line path in the $(\mu, R)$ plane, while the shaded bands
  indicate the corresponding moving standard deviations. The
  sequential succession from oscillatory to chaotic to rotational
  motion is clearly visible as the parameters increase along the
  ridge.}\label{fig:fractions_curve}
\end{figure}

To further investigate this order-chaos-order transition, we evaluated
the trajectory fractions along the curve of maximum chaotic abundance
defined by Eq.~\eqref{eq:maximums}, 
as shown in Fig.~\ref{fig:fractions_curve}. Along
this path, the chaotic
fraction is negligible for small parameter values, reaches a global
maximum at intermediate values (specifically around the autoparametric
resonance at $\mu = 4$) and then levels off for larger
parameters. Conversely, the fraction of oscillating
trajectories is almost one at low energies but decreases sharply as
chaos emerges. Meanwhile, the fraction of rotating trajectories,
initially zero, grows rapidly along the curve once the system acquires
sufficient energy to complete full revolutions,  as illustrated in
Fig.~\ref{fig:fractions_curve}.

\begin{figure}[ht]
\centering
\includegraphics[width=0.750\textwidth]{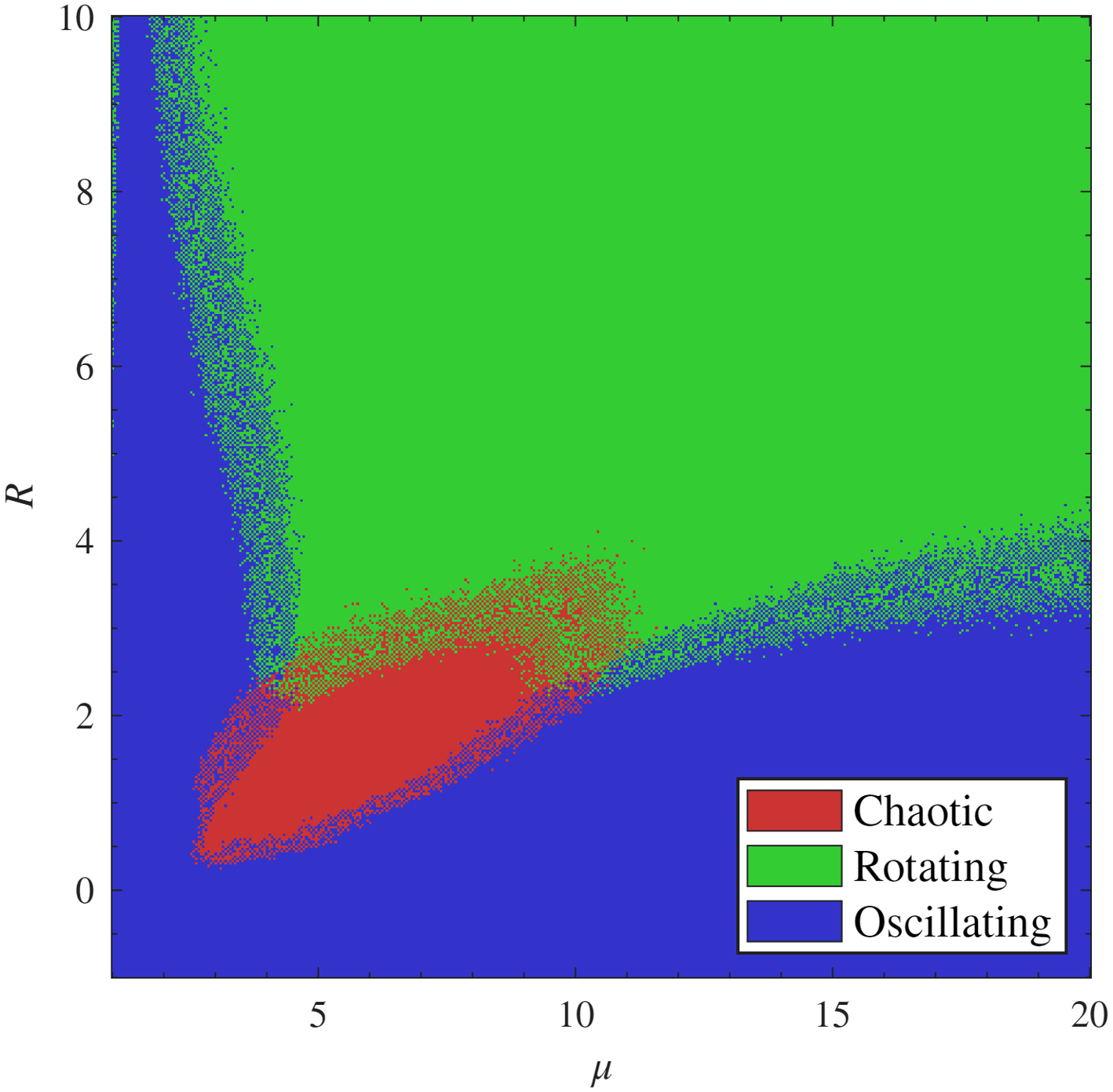}
\caption{Most abundant trajectory type at each point of the $(\mu, R)$ 
parameter plane, color-coded as follows: oscillatory (blue), rotational 
(green), and chaotic (red). Black points indicate regions where the most 
abundant type exceeds the second most abundant by less than $5\%$, 
signaling transition zones where no single type holds a clear majority. }\label{fig:dominating}
\end{figure}

This global phase-space organization is summarized in 
Fig.~\ref{fig:dominating}, which displays, for each point $(\mu, R)$ 
of the parameter plane, the most abundant trajectory type using a 
distinct color: blue for oscillatory, green for rotational, and red 
for chaotic. The 
diagram clearly delineates three distinct regimes: oscillatory motion 
predominates at low values of both $\mu$ and $R$, chaotic trajectories 
form a bounded central cloud at intermediate values, and rotational 
orbits prevail for sufficiently large values of $\mu$ and $R$. 
Crucially, the order-chaos-order transition is not a global property 
of the parameter plane but occurs specifically along paths that 
traverse the chaotic cloud, passing through the oscillatory-dominated 
region at low energies, across the chaotic cloud at intermediate 
values, and into the rotational-dominated region at higher energies.

\subsection{Energy exchanges and mode coupling}
\label{sec:energy}
To further characterize the dynamical regimes identified in the previous 
section, we analyze the internal energy distribution and the power exchanged 
between the elastic, pendular, and coupling degrees of freedom, following the 
decomposition framework of De Souza et al.~\cite{de2018energy,de2022internal}.

\begin{figure}[ht]
\centering
\includegraphics[width=0.99\textwidth]{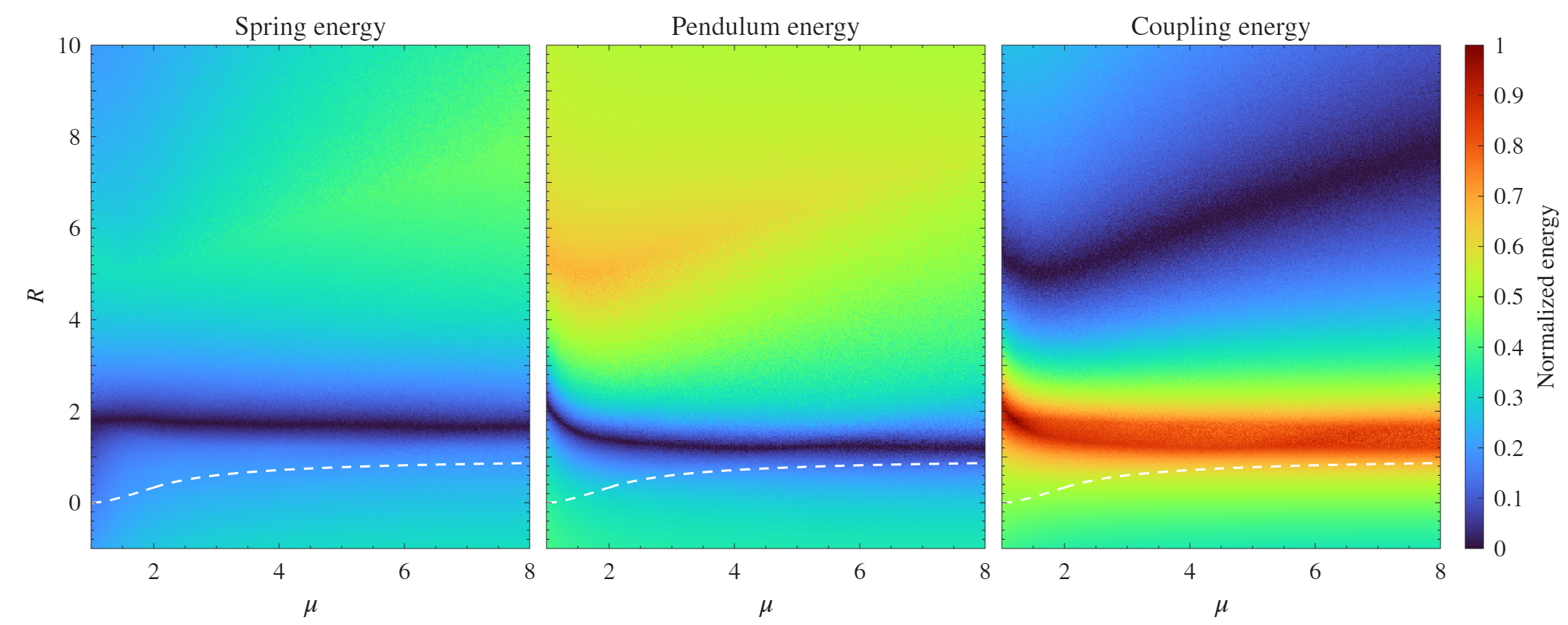}
\caption{Distribution of the normalized energy components across the $(\mu, R)$ parameter space. The panels respectively display the relative contributions of the spring energy, pendulum energy, and coupling energy to the total system energy. The color scale maps the normalized energy fraction from $0$ to $1$, with the dashed white line indicating the minimum energy for rotations to occur.}\label{fig:energy}
\end{figure}

Figure~\ref{fig:energy} shows the normalized energy components 
mapped over the $(\mu, R)$ parameter space. The three panels display the 
relative contributions of the spring energy, pendulum energy, and coupling 
energy to the total system energy, respectively. A clear hierarchy in the energy exchange rates emerges from these phase-space structures. At low values of $R$, the 
spring energy dominates, consistent with the prevalence of oscillatory 
trajectories in this region. As $R$ increases, the pendulum energy grows 
and the coupling energy develops a prominent peak concentrated along a 
well-defined band in parameter space, which closely coincides with the 
chaotic cloud identified in Fig.~\ref{fig:dominating}. The dashed white 
line marks the minimum energy required for rotational orbits to exist, 
above which the pendulum energy fraction increases sharply.

\begin{figure}[ht]
\centering
\includegraphics[width=0.85\textwidth]{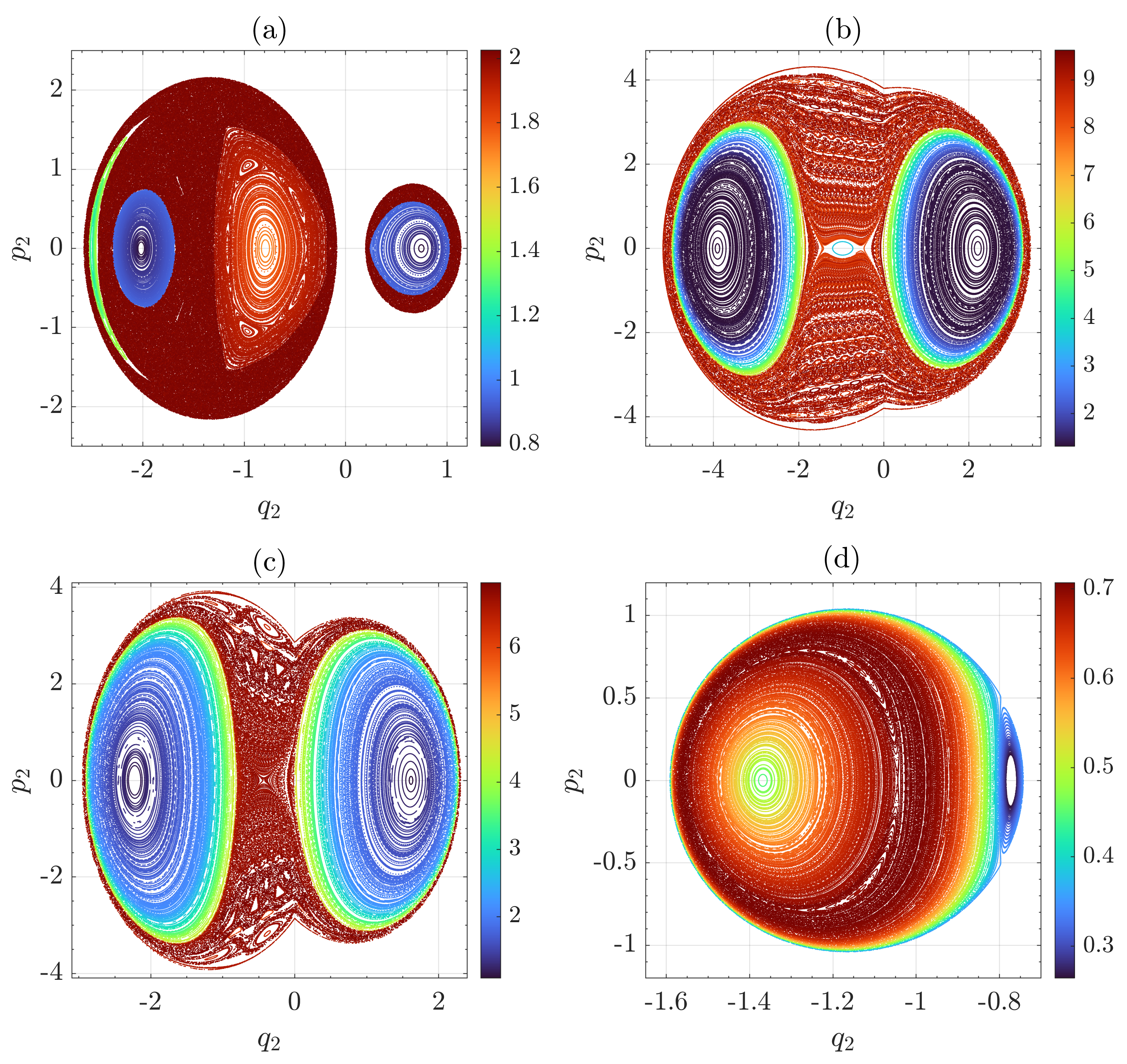}
\caption{Poincaré sections at the plane $q_1 = 0$ with $p_1 > 0$, the color corresponds to each trajectory $P_{max}$. The panels correspond to $(\mu, R)$ = (a) $(4, 1)$, (b) $(2.5, 6)$, (c) $(6, 6)$, and (d) $(7, -0.5)$. The color bar indicates the highest energy transfer rate (maximum absolute derivative of the energy components) reached along each individual trajectory, revealing how dynamic intensity varies across the regular and chaotic regions of the phase space.}\label{fig:poincare_power}
\end{figure}

A natural complement to the energy distribution is the rate at which 
energy is exchanged between the elastic, pendular, and coupling modes. 
We define the maximum power $P_{\max}$ as the maximum absolute value of 
the time derivative of a given energy component along a trajectory. 
Figure~\ref{fig:poincare_power} maps $P_{\max}$ for the coupling energy 
across the same Poincaré sections shown in Fig.~\ref{fig:psections}, 
revealing a clear hierarchy in the energy exchange rates. Rotational 
trajectories consistently exhibit the lowest values of $P_{\max}$, 
while chaotic trajectories display the highest, indicating that chaotic 
motion drives the most intense instantaneous energy exchange through the 
coupling term. Oscillatory trajectories within the regular islands can 
also reach notably high coupling power values depending on the specific 
parameters, but their $P_{\max}$ remains slightly lower than that of 
the surrounding chaotic sea. Taken together, these results establish 
$P_{\max}$ as a reliable dynamical indicator that distinguishes between 
the three trajectory types and connects the geometry of phase space to 
the underlying mode-coupling activity.

\section{Conclusions}
\label{sec:con}

We have studied the phase-space organization of the planar elastic 
pendulum as a function of its two dimensionless control parameters, 
the reduced energy $R$ and the frequency ratio $\mu$, by systematically 
classifying trajectories as oscillatory, rotational, or chaotic across 
the $(\mu, R)$ parameter plane via Poincaré sections.

The chaotic fraction is not uniformly distributed across the parameter plane but concentrates in a well-defined central cloud whose ridge follows the linear relation $R_{\max} = 0.34\,\mu - 0.38$; at its most intense, chaotic motion becomes abundant but never fills the entire accessible phase space, a regular component always persisting.
Outside this cloud, regular trajectories predominate: oscillatory orbits 
dominate at low values of both $\mu$ and $R$, consistent with the 
near-integrable behavior expected in the small-amplitude limit, while 
rotational orbits prevail for sufficiently large values of both 
parameters. The order-chaos-order transition is therefore not a global 
property of the parameter plane but occurs specifically along paths that 
traverse the chaotic cloud. Along the ridge of this cloud, a clear 
sequential mechanism emerges: oscillatory orbits progressively give way 
to chaotic trajectories, which in turn yield to rotational orbits once 
the energy is sufficient for the mass to complete full revolutions. The 
onset of rotational motion is gradual rather than sharp, reflecting a 
strong dependence on initial conditions.

The internal energy decomposition into spring-like, pendulum-like, and 
coupling contributions reveals that the coupling energy and coupling 
power are markedly enhanced precisely where chaotic trajectories 
predominate, establishing a direct link between inter-mode energy 
exchange and dynamical complexity. In the regular regions flanking the 
chaotic cloud, one mode largely decouples from the other and the motion 
recovers a near-integrable character.

These findings build on and complement the results of de Souza et al.~\cite{de2018energy,de2022internal}, who already examined part of the control-parameter space, by extending the analysis to the entire $(\mu, R)$ plane, confirming that enhanced energy exchange is a reliable indicator of chaotic dynamics, and clarify the structure of the chaotic cloud identified by Van der Weele and de Kleine~\cite{van1996order} by connecting it to the underlying mode-coupling mechanisms.

Overall, the main contribution of this work is a global and quantitative characterization of the dynamics over the full $(\mu, R)$ plane, extending beyond the individual trajectories and partial parameter ranges considered in earlier studies. Within this framework we have characterized the chaotic cloud and the linear law obeyed by its ridge, shown that chaos, although abundant, never fully dominates the accessible phase space, and ---most importantly from a physical standpoint--- identified inter-mode energy exchange as the mechanism organizing the order--chaos--order transition, thereby providing a transparent and quantitative indicator of chaos applicable across the entire parameter plane.
Future work will address the fractal structure of the boundaries between
dynamical regimes, the role of specific resonances, in particular the
autoparametric resonance at $\mu = 4$, in organizing the chaotic
cloud, and the extension of this analysis to the three-dimensional 
(non-planar) elastic pendulum.
\backmatter



\bmhead{Acknowledgements}

The authors gratefully acknowledge financial support from Project CISC 
I+D: \textit{Predictabilidad, caos, regularidad y simetrías en sistemas  físicos no lineales} (Project No.~22520240100022UD), funded by~CSIC- UdelaR. E.D.L. acknowledges support from Brazilian agencies CNPq (No. 304398/2023-3) and FAPESP (No. 2021/09519-5 and 2025/27957-0).

\section*{Data availability statement}
The Julia scripts used to generate all the data and figures presented in this
article are openly available in the Zenodo repository at
\url{https://doi.org/10.5281/zenodo.21876828}. Any further information is
available from the corresponding author upon reasonable request.

\section*{Declarations}

Conﬂict of interest The authors declare no competing
interests.









\bibliography{chaos.bib}

\end{document}